\documentclass[twocolumn,showpacs,preprintnumbers,amsmath,amssymb,prl,longbibliography]{revtex4-1}

\usepackage{graphicx}
\usepackage{dcolumn}
\usepackage{bm}
\usepackage{color}

\preprint{APS/123-QED}

\begin{document}

\title{Dynamical current-induced ferromagnetic and antiferromagnetic resonances}
\author{F. S. M. Guimar\~aes$^{1}$, S. Lounis$^{1}$, A. T. Costa$^{2}$, and R. B. Muniz$^{2}$}
\affiliation{$^1$ Peter Gr\"unberg Institut and Institute for Advanced Simulation, Forschungszentrum J\"ulich \& JARA, D-52428 J\"ulich, Germany\\ $^2$ Instituto de F\'{\i}sica, Universidade Federal Fluminense, Niter\'oi, Brazil}

\date{\today}

\begin{abstract}

We demonstrate that ferromagnetic and antiferromagnetic excitations can be triggered by the dynamical spin accumulations induced by the bulk and surface contributions of the spin Hall effect. Due to the spin-orbit interaction, a time-dependent spin density is generated by an oscillatory electric field applied parallel to the atomic planes of Fe/W(110) multilayers. For symmetric trilayers of Fe/W/Fe in which the Fe layers are ferromagnetically coupled, we demonstrate that only the collective out-of-phase precession mode is excited, while the uniform (in-phase) mode remains silent. When they are antiferromagnetically coupled, the oscillatory electric field sets the Fe magnetizations into elliptical precession motions with opposite angular velocities. The manipulation of different collective spin-wave dynamical modes through the engineering of the multilayers and their thicknesses may be used to develop ultrafast spintronics devices. Our work provides a general framework that probes the realistic responses of materials in the time or frequency domain. 

\end{abstract}

\maketitle

The interplay between charge, spin and orbital angular momentum in nano-structured systems is significantly widening the prospects of future technologies \cite{Bader:2010gr,Zutic:2004fo}. Spin-orbit coupling (SOC) is responsible for a variety of fascinating phenomena in condensed matter physics. For example, the lack of inversion symmetry activates the Dzyaloshinskii-Moriya interaction which favors the occurrence of non collinear ground-state magnetic configurations \cite{Bode:2007em,Menzel:2012bt,BezerraNeto:2013by}. Combined with time-reversal symmetry, it leads to protected conducting states in the so-called topological insulators \cite{Hasan:2010ku}, where spin injection and spin-to-charge conversion were recently demonstrated with the spin-pumping technique \cite{Shiomi:2014jy}. In fact, the generation of spin currents and spin accumulations by an electric current, in particular, has been a subject of much interest and research recently \cite{Avci:2015jp,Olejnik:2015bc,Lee:2015hk,Garello:2013fa,Liu:2012co,Kimura:2007fc,Valenzuela:2006cs,Kato:2004ft,Sanchez:2013kb}. Several groups showed that these non-equilibrium quantities can be used to set a magnetization into precessional motion in metallic systems \cite{Zhang:2015ky,Liu:2011fp,Fang:2011ch}, including antiferromagnets \cite{Zhang:2014ke}. Two recent reviews of the major experimental and theoretical results concerning the charge-to-spin conversion are outlined in Refs. \cite{Jungwirth:2012em,Sinova:2014wr}, for both metal and semiconductor devices. 

So far, theoretical approaches to current-induced spin currents, accumulations and torques in systems with more elaborate electronic structures are restricted to the case in which the applied electric field is static \cite{Freimuth:2014kq,Zelezny:2014ja,Tokatly:2015eu,Tanaka:2008jv, Guo:2008bx}. Here, we take it one step further, and investigate the dynamic magnetic response which is driven by a time-dependent electric field, as realized in the original experiments reported in Refs. \cite{Liu:2011fp,Kondou:2012dl,Fang:2011ch, Zhang:2015ky}. One advantage of such an electronic-structure-based method is that it naturally includes all surfaces, interfaces, and bulk contributions \cite{Haney:2013bp,Borge:2014fd,Lee:2015hk} to the spin Hall effect, including the coupling between local moments and the current-induced spin accumulation of conduction electrons \cite{Zhang:2002bd}, the transparency through the interface \cite{Zhang:2015ky}, and the spin-dependent scattering by the surfaces and interfaces \cite{Tokatly:2015eu}. Our framework is general enough to describe all kinds of dynamical Hall effects (which may be called ac Hall effects) and their reciprocal counterparts. We focus here, however, on the intrinsic (band-related) contributions to the ac spin Hall effect only.

In this Rapid Communication, we shall develop a microscopic theory for the current-induced magnetic response based on the premise that the amplitude of the external electric field is sufficiently weak to allow us to explore its effects within linear response theory. In this framework, we demonstrate---in ultrathin films of Fe and W(110)---that ferromagnetic resonances can be induced by ac electric fields owing to the spin-orbit interaction, and distinct modes can be excited depending on the type of magnetic interaction between the magnetic layers (ferromagnetic or antiferromagnetic). Implicitly, the excitation of the spin-wave modes indicates the presence of spin-orbit torques that are dynamical in nature. The studied phenomena are the reciprocal of the ac spin pumping and inverse spin Hall effect, which are one order of magnitude larger than their dc counterpart \cite{Wei:2014bn}---which adds up to the importance of a dynamical description. The considered applied electric field couples to the charge density, and we are able to calculate the induced spin disturbances and spin currents along the transverse directions of the external field, up to first order in the field intensity. We show that these quantities can be expressed in terms of generalized susceptibilities that may be calculated with the use of the random-phase approximation (RPA) of many-body theory. The additional complexity that arises when the RPA decoupling scheme is carried out in the presence of the spin-orbit interaction is the appearance of four coupled equations involving four distinct response functions that must be solved simultaneously \cite{Costa2010}. 

Here we are mainly interested in systems based on transition metals where Coulomb interactions play an important role. Thus, to accomplish this task explicitly, we consider that the electronic structure is described quite generally by a Hamiltonian $\hat{H} = \hat{H}_\text{0}+\hat{H}_\text{int}+\hat{H}_ \text{so}$, where $\hat{H}_\text{0}$ symbolizes the electronic kinetic energy plus a spin-independent local potential, $\hat{H}_\text{int}$ denotes the electron-electron interaction, and $\hat{H}_\text{so}$ stands for the spin-orbit interaction term. We choose an atomic basis set to represent these operators, which then acquire the following forms, $\hat{H}_\text{0} = \sum_{ij \sigma} \sum_{\mu\nu} t_{ij}^{\mu\nu} c_{i\mu \sigma}^{\dag}c_{j\nu \sigma}$, where $c^{\dagger}_{i\mu\sigma}$ creates an electron of spin $\sigma$ in atomic orbital $\mu$ on the site at ${\bf R}_i$, and the transfer integrals $t_{ij}^{\mu\nu}$ are parametrized following the standard Slater-Koster tight-binding formalism \cite{Slater:1954hi}. We assume that the effective electron-electron interaction $U$ is of short range, and keep only on-site interactions in $\hat{H}_\text{int}$. Hence, $\hat{H}_\text{int}=\frac{1}{2}\sum_{i \mu\nu}\sum_{\mu'\nu'}\sum_{\sigma\sigma'} U_{i;\mu\nu,\mu'\nu'}c^{\dagger}_{i\mu\sigma}c^{\dagger}_{i\nu\sigma'}c_{i\nu'\sigma'}c_{i\mu'\sigma}$, where $U_{i;\mu\nu,\mu'\nu'}$ is a matrix element of the effective electron interaction between orbitals, all centered on the same site $i$. In the spin-orbit term we also take into account intra-atomic interactions only, and write $\hat{H}_\text{so} = \sum_{i\mu\nu} \sum_{\sigma \sigma'} \xi_i \langle i\mu\sigma |{\bf L}\cdot {\bf S} |i\nu \sigma' \rangle c_{i\mu \sigma}^{\dag}c_{i\nu \sigma'}$, where $\xi_i$ denotes the spin-orbit coupling constant on site $i$, and ${\bf L}$ and {\bf S} are the orbital angular momentum and spin operators, respectively. 

In order to calculate the desired spin responses in the presence of the spin-orbit interaction, it is useful to introduce the generalized spin susceptibilities 
\begin{equation}\label{suscgeneric}
\chi^{{\sigma_1\sigma_2\sigma_3\sigma_4}\mu\nu\gamma\xi}_{ijk\ell}(t) = -\frac{i}{\hbar}\Theta(t)\langle[c^{\dag}_{i\mu\sigma_1}(t) c_{j\nu\sigma_2}(t),c^{\dag}_{k\gamma\sigma_3} c_{\ell\xi\sigma_4}] \rangle\ ,
\end{equation}
where each $\sigma_i$ symbolizes either $\uparrow$ or $\downarrow$ spin directions. We may represent them as a 4$\times$4 matrix structure in spin space, whose rows and columns are labeled by pairs of spin indices $\sigma\sigma' = \uparrow\downarrow,\uparrow\uparrow,\downarrow\downarrow,\downarrow\uparrow$ ($+,\uparrow,\downarrow,-$). Within the RPA it is possible to express all elements in terms of the noninteracting spin susceptibilities $\chi^{(0)}$, that are generated by evaluating the commutators which enter into Eq. (\ref{suscgeneric}) in the noninteracting ground state. In matrix form the relation is schematically given by $\left[\chi(\omega)\right] = \left[\chi^{(0)}(\omega)\right]-\left[\chi^{(0)}(\omega)\right]\, \left[U\right]\left[\chi(\omega)\right]$, where 
\begin{equation}
\begin{split}
\chi^{(0)}_{ijk\ell}(\omega) =\hbar\int d\omega' f(\omega')&\left\{g_{jk}(\omega'+\omega)\Im [ g_{\ell i}(\omega') ]\right. \\
&+ \left. g^{-}_{\ell i}(\omega'-\omega)\Im [ g_{jk}(\omega') ]\right\}.
\end{split}
\end{equation}
Here, to simplify the notation, we have omitted the spin and orbital indices, assuming that they are included in the site indices. We define $\Im \left[g\right] =\frac{i}{2\pi}\left[g - g^{-}\right]$, where $g$ and $g^{-}$ represent the retarded and advanced one-electron propagators, respectively, and $f(\omega)$ is the usual Fermi distribution function. We remark that at this stage we are ignoring long-range Coulomb interactions which are relevant to ensure charge conservation, especially in the static limit of homogeneous fields. Edwards \cite{Edwards:2015tn} has recently shown that for bulk systems this may not be so significant for relatively small SOC. 

We begin by examining an ultrathin film of W(110) with atomic planes stacked along the $\mathbf{\hat{z}}$ direction, choosing the $\mathbf{\hat{x}}$ and $\mathbf{\hat{y}}$ Cartesian axes parallel to the layers, in the $[1 \bar{1} 0]$ and $[0 0 1]$ directions, respectively. Assuming $U=1$ eV and $\xi=0.26$ eV for W, and adjusting the center of its $d$ bands to reproduce the electronic occupations obtained by density functional theory (DFT) calculations \cite{aklautaupc} for each atomic plane, one finds that the ground state of the W film is nonmagnetic, as expected. Let us then suppose that a spatially uniform harmonic electric field $\boldsymbol{\mathcal{E}} = E_0 \cos(\omega t)\, \mathbf{\hat{u}}_\mathcal{E}$ is applied parallel to the layers in an arbitrary direction $\mathbf{\hat{u}}_\mathcal{E}$. In this case, the time-dependent perturbing Hamiltonian is given by
\begin{equation}\label{haetbml}
\begin{split}
\hat{V}(t)= \frac{eE_0}{\hbar\omega} \frac{1}{N}\sum_{\mathbf{k}_\|,\sigma}\sum_{\substack{\ell\ell'\\\mu\nu}}&\boldsymbol{\nabla}_{\mathbf{k_\|}} t_{\ell\ell'}^{\mu\nu}(\mathbf{k}_\|)\cdot\mathbf{\hat{u}}_\mathcal{E}\ \sin(\omega t)\\
&\times c^\dag_{\ell\mu\sigma}(\mathbf{k}_\|,t)c_{\ell'\nu\sigma}(\mathbf{k}_\|,t)\ ,
\end{split}
\end{equation}
where $\ell$ and $\ell'$ identify atomic planes, and $\bf{k}_{\parallel}$ is a wave vector parallel to the layer, belonging to the two-dimensional Brillouin zone. With the use of linear response theory we may calculate the components of the local spin disturbance per atom in plane $\ell_1$, induced by the ac applied electric field by virtue of the SOC. They are given by 
\begin{equation}
\begin{split}
\delta\langle \hat{S}_{\ell_1}^m(t)\rangle =& \mathcal{A}^m_{\ell_1}(\omega)\sin\left(\omega t-\phi^m_{\ell_1}(\omega)\right)
\end{split}\ ,
\end{equation}
where $\mathcal{A}^m_{\ell_1}(\omega) = \frac{eE_0}{\hbar\omega}|\mathcal{D}^m_{\ell_1}(\omega)|$ represents the amplitude of the local spin disturbance, and $\phi^m_{\ell_1}(\omega)$ is the frequency-dependent phase of the complex number
\begin{equation}
\begin{split}
\mathcal{D}^m_{\ell_1}(\omega)=& \sum_{\substack{\mathbf{k}_\|\\\sigma}}\sum_{\substack{\ell\ell'\\\mu\gamma\xi}}\chi^{^{m\sigma}\mu\mu\gamma\xi}_{\ell_1\ell_1\ell\ell'}(\mathbf{k}_\|,\omega)\boldsymbol{\nabla}_{\mathbf{k_\|}} t_{\ell\ell'}^{\gamma\xi}(\mathbf{k}_\|)\cdot\mathbf{\hat{u}}_\mathcal{E}\ .
\end{split}
\end{equation}
Here, $m=x, y, z$ labels the corresponding spin components, $\chi^{x \sigma} = [\chi^{\uparrow \downarrow \sigma \sigma} + \chi^{\downarrow \uparrow \sigma \sigma}]/2$, $\chi^{y \sigma} = [\chi^{\uparrow \downarrow \sigma \sigma} - \chi^{\downarrow \uparrow \sigma \sigma}]/2i$, and $\chi^{z \sigma} = \chi^{\uparrow \uparrow \sigma \sigma} - \chi^{\downarrow \downarrow \sigma \sigma}$.

\begin{figure}
\includegraphics[width = 8.6cm]{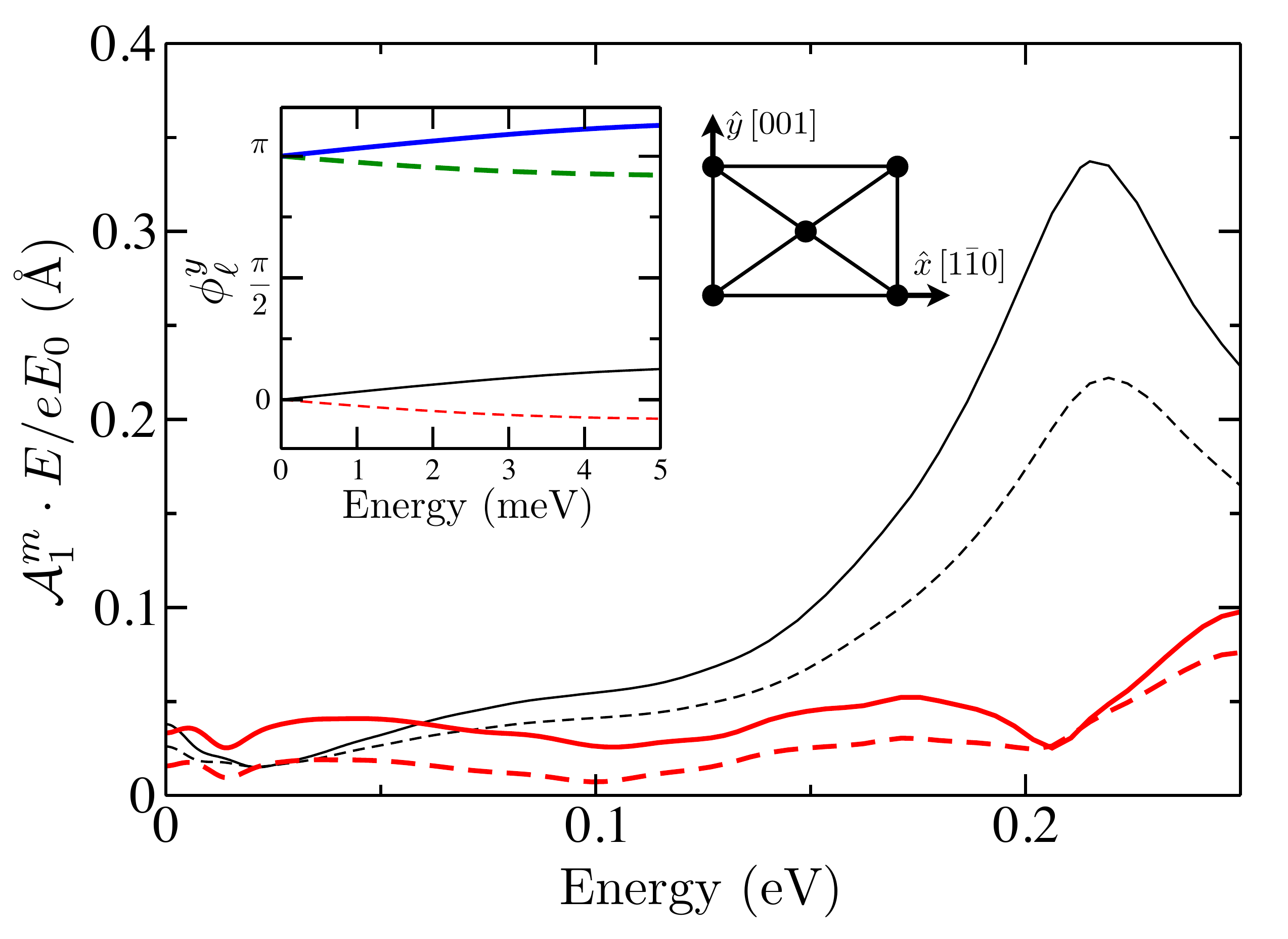}
\caption{(Color online) Amplitudes of the surface spin disturbances $\mathcal{A}_1^y(E)$ (thin black lines) and $\mathcal{A}_1^x(E)$ (thick red lines) induced in a free-standing slab of W(110) by ac electric fields applied along the $[1 \bar{1} 0]$ and $[0 0 1]$ directions, respectively. The slab comprises four atomic planes which are labeled sequentially by $\ell = 1$--$4$, starting from one of the W surfaces. Solid lines represent results calculated for $U= 1$ eV and and dashed lines for $U=0$. The inset shows the corresponding phases $\phi^y_{\ell}(E)$ calculated with $U=1$ eV for $\ell = 1$ (black thin solid line) $\ell = 2$ (red thin dashed line), $\ell = 3$ (green thick dashed line), and $\ell = 4$ (blue thin solid line), as a result of an electric field applied along $[1 \bar{1} 0]$.}
\label{W4}
\end{figure}

Due to the presence of SOC, an ac electric field applied along the $[1 \bar{1} 0]$ ($\mathbf{\hat{x}}$) direction should produce an ac spin accumulation $\langle \hat{S}_{\ell}^y\rangle \ne 0$ in the W(110) atomic planes as a result of the bulk spin currents generated by the dynamic spin Hall effect and also from the spin-orbit fields originated in the spin-split surface states. It also gives rise to a bulk pure ac spin current with spin polarization $\hat{z}$ that flows parallel to the layer along the $[0 0 1]$ ($\mathbf{\hat{y}}$) direction, but leads to no spin accumulation due to the translation symmetry of the layers. Similarly, if the field is applied along the $[0 0 1]$ direction, the W(110) atomic planes are expected to acquire an ac spin accumulation $\langle \hat{S}_{\ell}^x\rangle \ne 0$. In this case, the electric field also generates an ac spin current with spin polarization $\hat{z}$ that flows along the $[1 \bar{1} 0]$ direction, causing no spin accumulation. This is precisely what we have found in our calculations of the spin disturbances and currents induced in a free-standing slab of W(110). The results for the amplitudes and phases of $\delta\langle \hat{S}_{\ell}^m(t)\rangle$ calculated as functions of the energy $E=\hbar\omega$ are shown in Fig. $\ref{W4}$ for electric fields applied in two perpendicular directions. Owing to the spatial anisotropy of the (110) two-dimensional lattice, the amplitudes of the spin accumulation in the W surface differ considerably for electric fields applied along the $[1 \bar{1} 0]$ and $[001]$ directions. One can also appreciate the importance of the Coulomb exchange interaction within the W layer by comparing the amplitudes of the induced magnetic moments obtained with $U=1$ eV and $U=0$, which are depicted by the solid and dashed lines, respectively, in Fig. $\ref{W4}$. The overall increase for $U\ne 0$ suggests that these effects possibly may be used to excite spin fluctuations (paramagnons) in ultrathin films of nearly ferromagnetic metals such as Pd and Pt, which exhibit relatively large Stoner enhancement factors. The inset illustrates the corresponding phases $\phi^m_{\ell}(E)$ of the spin disturbances induced in the four W atomic planes by an electric field applied along $[1 \bar{1} 0]$ with $U=1$ eV. For low values of $\omega$ we identify a current-induced staggered spin disturbance profile on the W(110) atomic planes. The same feature appears when the field is applied along the $[0 0 1]$ direction for both values of $U$. This is compatible with the charge current leading to spin accumulations of inverse sign on the opposite W surfaces, and the spin polarization induced by this spin imbalance in each surface decreases as one moves into the W film along the stacking direction in an oscillatory manner with a period of approximately two inter-planar distances, thus favoring the antiferromagnetic alignment. 

\begin{figure}
\includegraphics[width = 8.6cm]{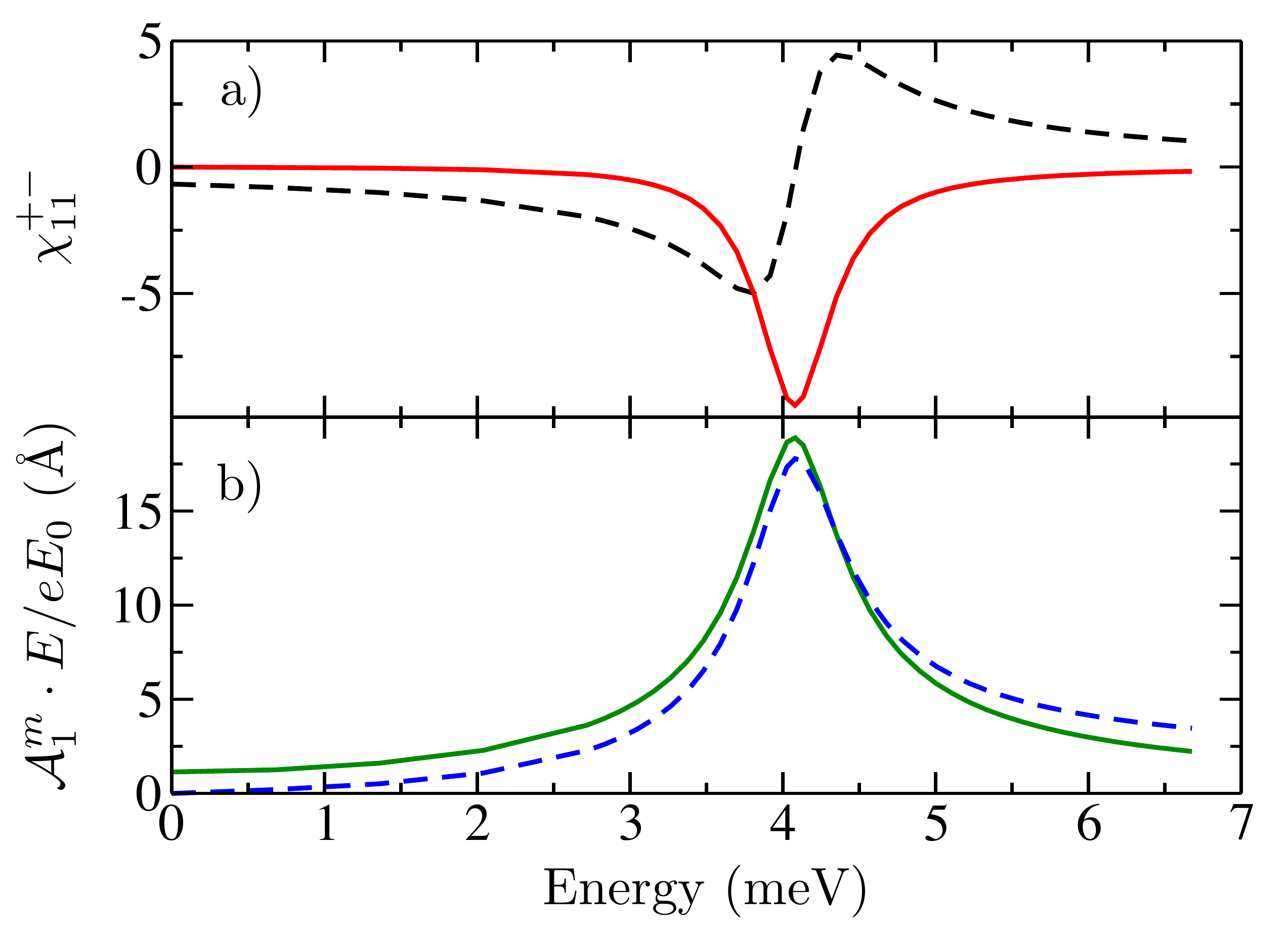}
\caption{(Color online) (a) Real (black dashed line) and imaginary (red solid line) parts of the local transverse spin susceptibility calculated (in arbitrary units) for a monolayer of Fe/W(110) as functions of energy. (b) Amplitudes of the local induced spin disturbances $\mathcal{A}^y(E)$ (green solid line) and $\mathcal{A}^z(E)$ (blue dashed line) calculated in the Fe surface layer.}
\label{FeW}
\end{figure}

We shall now discuss the use of the ac charge current as a way of exciting spin-wave modes in an Fe layer adsorbed to a thin film of W(110), consisting of five atomic planes in total. The ground-state magnetization of the Fe layer in this case sets down in-plane along the $[1 \bar{1} 0]$ direction, which is the easy axis. The uniform spin-wave mode observed in a ferromagnetic resonance (FMR) absorption spectrum is revealed as a resonance in the transverse dynamical spin susceptibility, which represents the response of the system to a time-dependent oscillatory transverse magnetic field. This is clearly shown in Fig. $\ref{FeW}$(a), which depicts the local transverse spin susceptibility $\chi^{+-}_{11}(q_\parallel = 0,E)$ calculated as a function of energy $E=\hbar \omega$ in the Fe surface layer. The peak position in Im $\chi^{+-}_{11}(E)$ is the anisotropy energy due to the spin-orbit interaction, and the linewidth of the resonance is inversely proportional to the spin-wave lifetime. If instead of a transverse magnetic field we apply an oscillatory electric field along the easy-axis direction, for example, we may also calculate  the current-induced spin disturbances in the Fe layer $\delta\langle \hat{S}_{1}^m(t)\rangle$ within our approach, and their calculated amplitudes $\mathcal{A}_1^m (E)$ are illustrated in Fig. $\ref{FeW}$(b). They clearly show that both transverse components of the induced spin disturbances in the Fe layer exhibit a peak precisely at the ferromagnetic resonance energy, demonstrating that the oscillatory electric field is exciting the uniform spin-wave mode by means of the dynamical spin-orbit torque. We see the appearance of an oscillatory spin disturbance $\delta\langle \hat{S}_{1}^z(t)\rangle$, with polarization perpendicular to the Fe surface layer, which is dephased by approximately $\pi/2$ from $\delta\langle \hat{S}_1^y(t)\rangle$, revealing that the magnetization of the Fe layer is set into an elliptic precessional motion around the easy axis. We note that the $y$ $(z)$ component is even (odd) with respect to magnetization inversion ($\mathbf{M}\rightarrow-\mathbf{M}$), as discussed in Ref. \cite{Freimuth:2014kq}. We have also calculated the change in orbital angular momentum induced in the Fe surface layer by the same electric field. Both amplitudes of $\delta\langle \hat{L}_1^y(t)\rangle$ and $\delta\langle \hat{L}_1^z(t)\rangle$ display well-defined maxima at the same ferromagnetic resonance energy, but they are approximately one order of magnitude smaller than the corresponding values for $\mathcal{A}_1^m (E)$.

\begin{figure}
\includegraphics[width = 8.6cm]{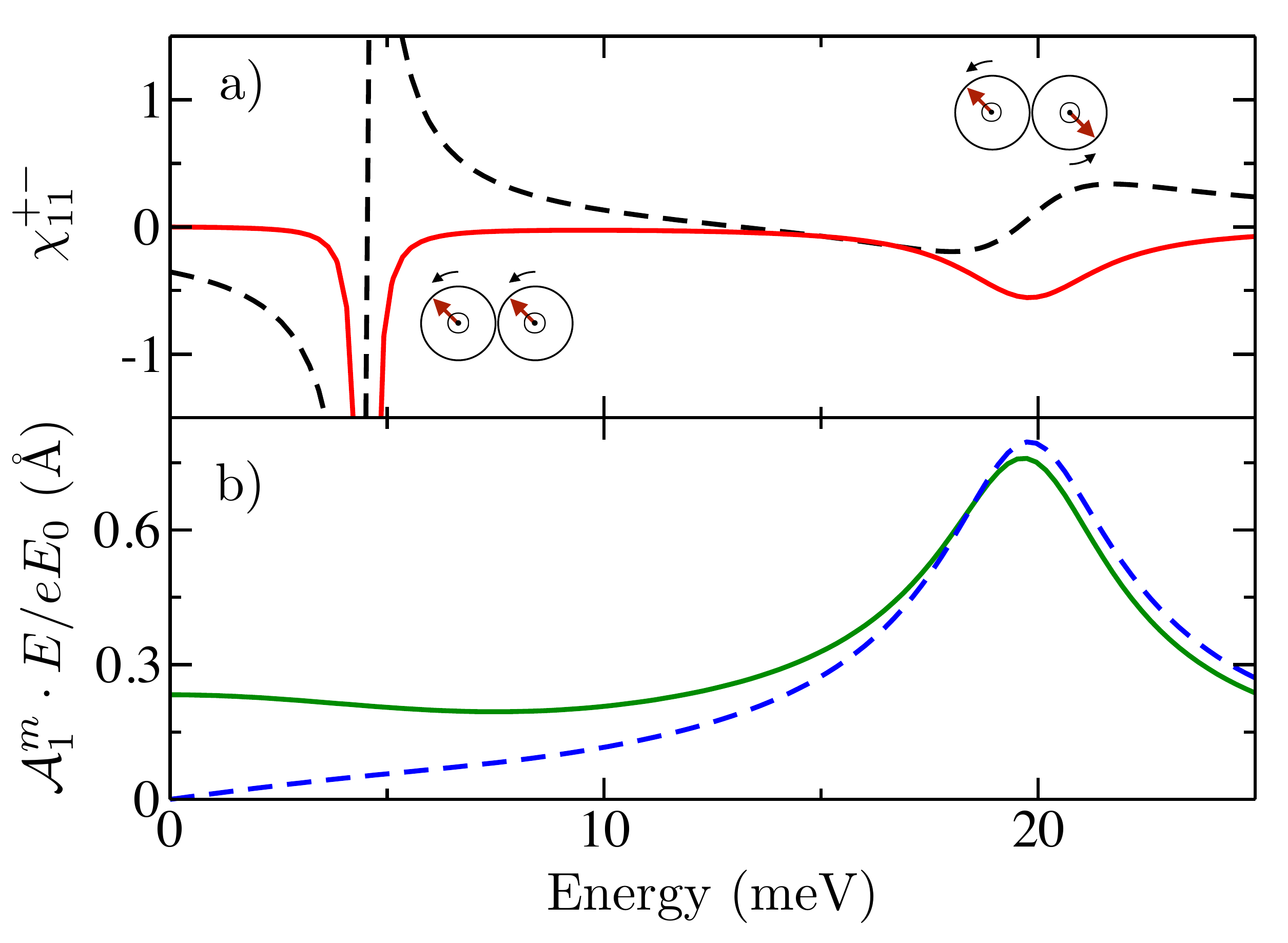}
\caption{(Color online) Same as in Fig \ref{FeW} for a Fe/W(110)/Fe trilayer. }\label{FeWFe}
\end{figure}

We now turn our attention to Fe/W(110)/Fe multilayers. We consider two different thicknesses for the tungsten spacer layer, starting with two atomic planes of W where the magnetizations of the Fe layers are ferromagnetically coupled along the long axis. In this situation, the FMR absorption spectrum exhibits two precession modes corresponding to the cases in which those magnetizations oscillate in phase (acoustic mode) and out of phase (optical mode), respectively. This is clearly visible in Fig. $\ref{FeWFe}$(a) which shows the local transverse spin susceptibility calculated as a function of energy for one of the Fe surface layers. The energy difference between the two peaks in Im $\chi^{+-}_{11}(E)$ is a measure of the exchange coupling between the Fe magnetizations. In Fig. $\ref{FeWFe}$(b) we present our calculated results for the amplitudes of the transverse spin components induced in the same Fe surface by an oscillatory electric field applied along the $[1 \bar{1} 0]$ direction. They show that only the out-of-phase precession mode is excited by the electric field, while the uniform (in-phase) precession mode remains silent. This is reasonable for a perfectly symmetric configuration such as the one we are considering, since the oscillatory spin accumulations that drive the magnetizations of the opposite Fe layers into precession are $180^{\circ}$  out of phase. Indeed, the phase differences $\phi^{y,z}_1(\omega) - \phi^{y,z}_4(\omega)$ between the spin disturbances induced in the Fe surfaces are both equal to $\pi$ for all values of $\omega$. This contrasts with traditional FMR experiments, driven by a time-dependent homogeneous transverse magnetic field, where the optical mode would not be observed, unless the individual FM layers have different resonance frequencies. Deposition of the layered structure on substrates introduces an asymmetry between the ferromagnetic layers that may prevent complete cancellation of the torques, enhancing the acoustic-mode signal. However, this can be tuned by a suitable choice of substrate.

\begin{figure}
\includegraphics[width = 8.6cm]{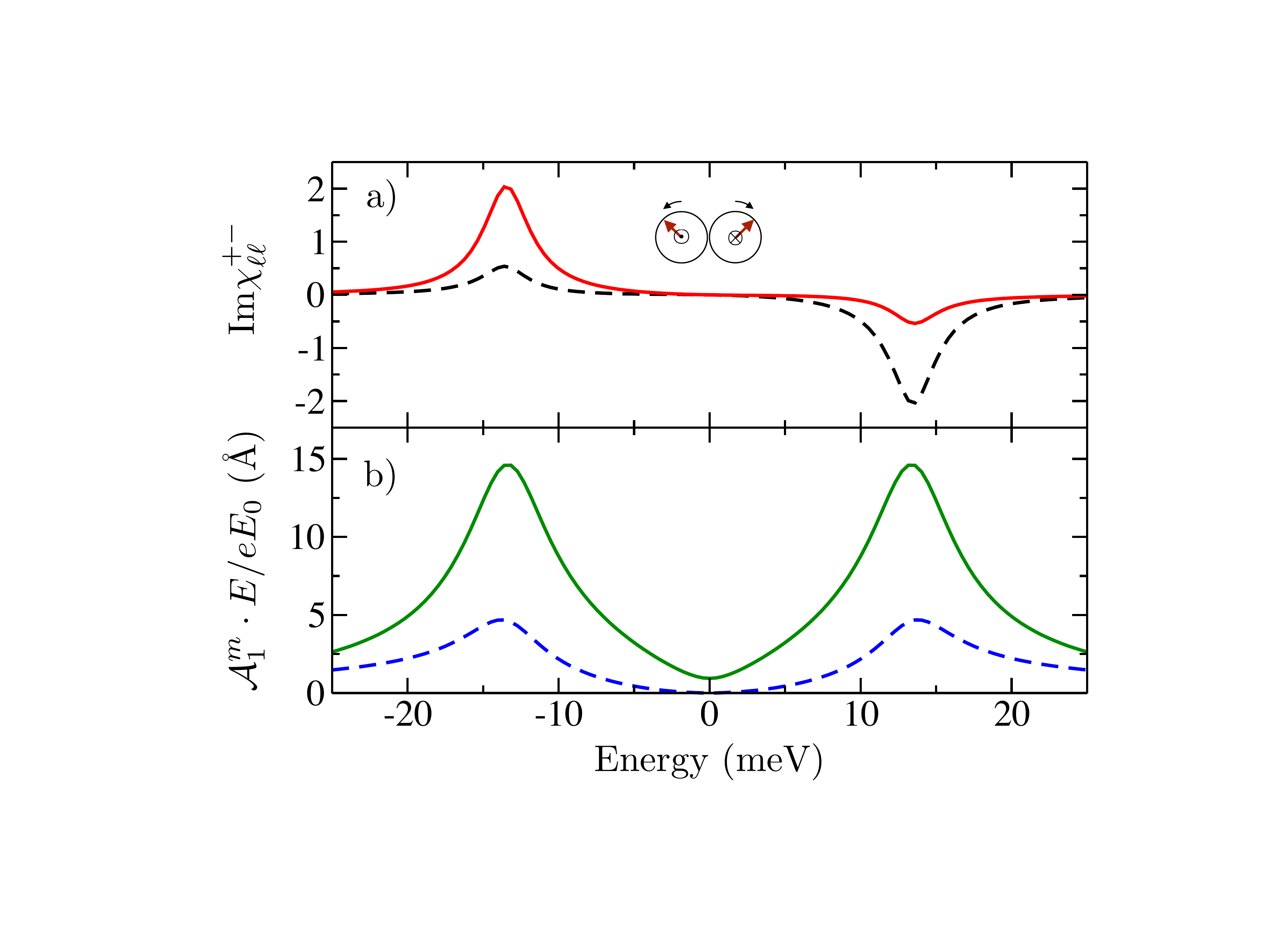}
\caption{(Color online) (a) Im$\chi^{+-}_{11}(E)$ (red solid line), and Im$\chi^{+-}_{55}(E)$ (black dashed line), calculated (in arbitrary units) as functions of energy for the Fe surface layers of the antiferromagnetically coupled Fe/W(110)/Fe trilayer. The W spacer layer has three atomic planes. (b) Amplitudes of the induced local-spin-disturbance components $\mathcal{A}_1^y(E)$ (green solid line) and $\mathcal{A}_1^z(E)$ (blue dashed line) calculated for one of the Fe surface layers. }\label{FeWFe_AF}
\end{figure}

By increasing the thickness of the W spacer layer to three atomic planes, we find that the magnetizations of the Fe layers become antiferromagnetically coupled. We label the two Fe surfaces in this trilayer by 1 and 5, respectively. In fact, assuming that in the ground state the Fe magnetizations are ferromagnetically aligned, a calculation of Im$\chi^{+-}_{11}(E=\hbar\omega)$ displays two resonant spin-wave modes---one at a positive angular frequency and another at a negative value of $\omega$--proving that the Fe layers are indeed antiferromagnetically coupled in this case. However, one may also calculate the local transverse spin susceptibilities from the antiferromagnetic (ground) state. The results for the imaginary parts of $\chi^{+-}_{11}$ and $\chi^{+-}_{55}$, calculated as functions of energy, are shown in Fig. $\ref{FeWFe_AF}$(a). Each shows two extrema with different intensities at $\pm \omega_0$, which is consistent with the antiferromagnetic coupling between the Fe layers in the presence of the anisotropy field due to the SOC. In Fig. $\ref{FeWFe_AF}$(b) we present results for the amplitudes of the local spin disturbances $\mathcal{A}_1^y(E)$ and $\mathcal{A}_1^z(E)$ in one of the Fe surface layers. We also found the phase differences between the spin disturbances induced in the two Fe surface layers to be $\phi^{y}_1(\omega) - \phi^{y}_5(\omega) = \pi$, and $\phi^{z}_1(\omega) - \phi^{z}_5(\omega) = 0$, for all values of $\omega$. This is consistent with the two magnetizations being set into elliptic precessional motions around their equilibrium directions, however, with opposite angular velocities. 

To estimate the charge-to-spin conversion we define a coefficient $\gamma (E) = |\mathcal{A}_1(E)|/|j^C (E)|$ \cite{Tokatly:2015eu}, given by the ratio between the amplitudes of the surface-induced spin accumulation and of the charge current density $|j^C(E)|$. In the energy range of interest, $|j^C (E)|\,E/E_0$ is approximately constant and, as a result, the curves representing $\mathcal{A}_1(E)\,E/eE_0$ are basically the same as $\gamma(E)$, except for a constant multiplicative factor. It follows that the charge-to-spin conversion at the resonance frequency is largely enhanced with respect to its values at very low frequencies.

To summarize, we investigated dynamical transport properties in the context of charge-to-spin conversion. For instance, we evaluate spin and orbital angular momentum accumulation induced by an ac charge current mediated by the spin-orbit interaction. We demonstrate that specific spin-precession modes can be excited in thin films depending on the magnetic nature of the nanostructures, which may assist their switching and offer a potentially useful tool for ac spintronic developments and nanotechnologies. In fact, it was recently shown that the spin-wave excitation is directly related to the switching rate. \cite{Spinelli:2014db} Our framework allows for the inspection of additional phenomena, such as the whole family of dynamical Hall effects and all their reciprocal counterparts.

\begin{acknowledgments}

We are grateful to A. B. Klautau for providing tight-binding parameters, M. dos Santos Dias and S. Bl\"ugel for discussions, and computing time on JUROPA supercomputer at J\"ulich Supercomputing Centre. We are thankful for the support from FAPERJ, CAPES, and CNPq (Brazil), Alexander von Humboldt Foundation (Germany), and HGF-YIG Programme VH-NG-717 (Functional nanoscale structure and probe simulation laboratory-Funsilab).

\end{acknowledgments}

\bibliography{SHE.bib}
\end{document}